\documentclass{acm_proc_article-sp}
\usepackage{graphics,color,epsfig,xspace,enumerate,multirow}
\newdef{definition}{Definition}
\newtheorem{theorem}{Theorem}
\newcommand{\be}{\begin{eqnarray}}
\newcommand{\ee}{\end{eqnarray}}

\newcommand{\rt}{$T_{r}$\xspace}
\newcommand{\m}{$M$\xspace}
\newcommand{\sub}{}

\begin{document}

\title{Design and Analysis of an Asynchronous Zero Collision MAC Protocol}
\numberofauthors{2}
\author{
\alignauthor
Jiwoong Lee\\
       \affaddr{Department of Electrical Engineering and Computer Sciences}\\
       \affaddr{University of California at Berkeley}\\
       \affaddr{Berkeley, California 94720}\\
       \email{porce@eecs.berkeley.edu}
\alignauthor
Jean Walrand\\
       \affaddr{Department of Electrical Engineering and Computer Sciences}\\
       \affaddr{University of California at Berkeley}\\
       \affaddr{Berkeley, California 94720}\\
       \email{wlr@eecs.berkeley.edu}
}

\maketitle
\begin{abstract}
This paper proposes and analyzes a distributed MAC protocol that achieves zero collision with no control message exchange nor synchronization. ZC (ZeroCollision) is neither reservation-based nor dynamic TDMA; the protocol supports variable-length packets and does not lose efficiency when some of the stations do not transmit. At the same time, ZC is not a CSMA; in its steady state, it is completely collision-free.

The stations transmit repeatedly in a round-robin order once the convergence state is reached. If some stations skip their turn, their transmissions are replaced by idle $20 \mu$-second mini-slots that enable the other stations to keep track of their order. To claim the right to transmit, a station selects one of the idle mini-slots in one round and transmits in that mini-slot in the next round. If multiple stations select the same mini-slot, they collide and repeat their random selection in a subsequent round.  A station loses its right to transmit after being idle in a given number of successive rounds. The number of transmissions per round can be adapted to the number of active stations in the network.

Because of its short medium access delay  and its efficiency, the protocol supports both real-time and elastic applications. The protocol allows for nodes leaving and joining the network; it can allocate more throughput to specific nodes (such as an access point). The protocol is robust against carrier sensing errors or clock drift.

While collision avoidance is guaranteed in a single collision domain, it is not the case in a multiple collision one. However, experiments show ZC supports a comparable amount of goodput to CSMA in a multiple collision domain environment.

The paper presents an analysis and extensive simulations of the protocol, confirming that ZC outperforms both CSMA and TDMA at high and low load.

\end{abstract}




\section{Introduction}
Proliferation of mobile data devices, incessant increase of Internet usage, and growing coexistence of delay sensitive and delay tolerant traffic demand improvements of wireless communication performance. This paper proposes a new MAC protocol that is very efficient and supports real-time and elastic applications.  The mechanism is distributed, which is desirable in terms of overhead, fault-resilience, and implementation complexity.

A majority of CSMA \cite{1975KLEINROCK1} type MAC protocols use a limited level of memory for congestion relaxation purpose by increasing backoff range on collisions, but the propagation is reset on any transmission success, which limits the memory of the system (see \cite{1997IEEE80211}).
On the other hand, some reservation-type MAC protocols use strong memory. For instance, a station might respect a series of periodic time slots reserved for others.
Due to the random channel fading or unpredictable hardware imperfections however, the station can lose track of synchronization, and possibly produce collisions. Therefore some self-stabilization process is indispensable for any robust distributed MAC protocol.

In this research, we propose a distributed, collision-free, self-stabilizing single channel MAC protocol requiring no control messages. We call it the ZeroCollision (ZC) protocol.
In the steady state, the stations select a round-robin order in which they transmit.  If a station has a variable length packet to send at its own turn, it accesses the medium during a variable duration of transmission slot. Otherwise, it skips its turn and others in the network see an empty mini-slot, which is fixed duration and much shorter than a typical transmission slot. The existence of an empty mini-slot or a transmission slot is a sufficient statistic so that stations can count the order of transmission for its own access. No packet decoding is necessary for this operation.
In the transient state, any collision-experiencing station chooses one of the empty mini-slots in the next round. Stations repeat the random mini-slot selection until they stop experiencing collisions; they keep their assigned order thereafter.  A station loses its assignment if it stops transmitting for a number of consecutive rounds.

%

The contribution of this research is as follows.
\begin{itemize}
\item ZC results in short access delays, both for low and high load. This characteristic contrasts with the fact that TDMA-based MAC protocols inevitably exhibit excessive access delays at low load and that so do random access based MAC protocols at high load. ZC exhibits superior performance over a wide range of active number of stations in the system.
\item ZC does not involve an exchange of control messages among adjacent nodes. Therefore the assignment process is simple and robust, and does not incur a significant control overhead. It does not require stations to decode any message for access decision purpose. So the gap between interference range and transmission range does not affect the correct operations of ZC. Since the interference range is three or more times larger than the transmission range, the working coverage of ZC is substantially larger than that of some protocols requiring message decoding for access control purpose. The only necessary information for a station to decide when to access the channel is the cumulative length of silence, where silence is technically defined by subthreshold signal-to-noise ratio.
\item ZC is self-stabilizing: the network, by following the ZC algorithm, recovers from dynamic events such as sudden node arrivals or departures, from erroneous events such as carrier sensing error, which are typically caused by channel fading or hardware defect, and from corresponding clock drifts. It is necessary that the process should stabilize fast enough, relative to the occurrence frequency of dynamic or erroneous events. In section \ref{s.analysis}, we show the expected stabilization time in the worst case is upper-bounded, and typically around 2 to 3 seconds. In the nominal case where each individual station maintains some information about the network, the system stabilizes even faster. In addition, while ZC works in a non-infrastrucure mode by design, we provide a simpler way for the power saving node to retrieve its access slot by listening to a beacon slot of the access point, if any, in the infrastructure mode operation.
\item ZC achieves a collision-free state in a single collision domain. Although it is not guaranteed in a multiple collision domain due to hidden/exposed node syndrome, experimental results show comparable performance.

\end{itemize}
Simulations in section \ref{s.experiment} consider a few realistic modeling assumptions including two-way propagation time, capture effect, carrier sensing error, and clock drift. We also consider the case where more number of stations are active than a defined network capacity. By simulation, we provide performance comparison between ZC, CSMA, and TDMA, and confirms that ZC inherits advantages both from CSMA and TDMA. We also show that ZC enables desirable coexistence of two representative traffic patterns - delay sensitive periodic traffic (voice over Internet protocol, VoIP \cite{2002VOIP}) and delay tolerant best effort traffic (web transactions). In section \ref{s.voipdelaycapacity} we show that ZC substantially increases the VoIP capacity web traffic is present.

\begin{figure*}[!t]
  \begin{center}
  \epsfig{file=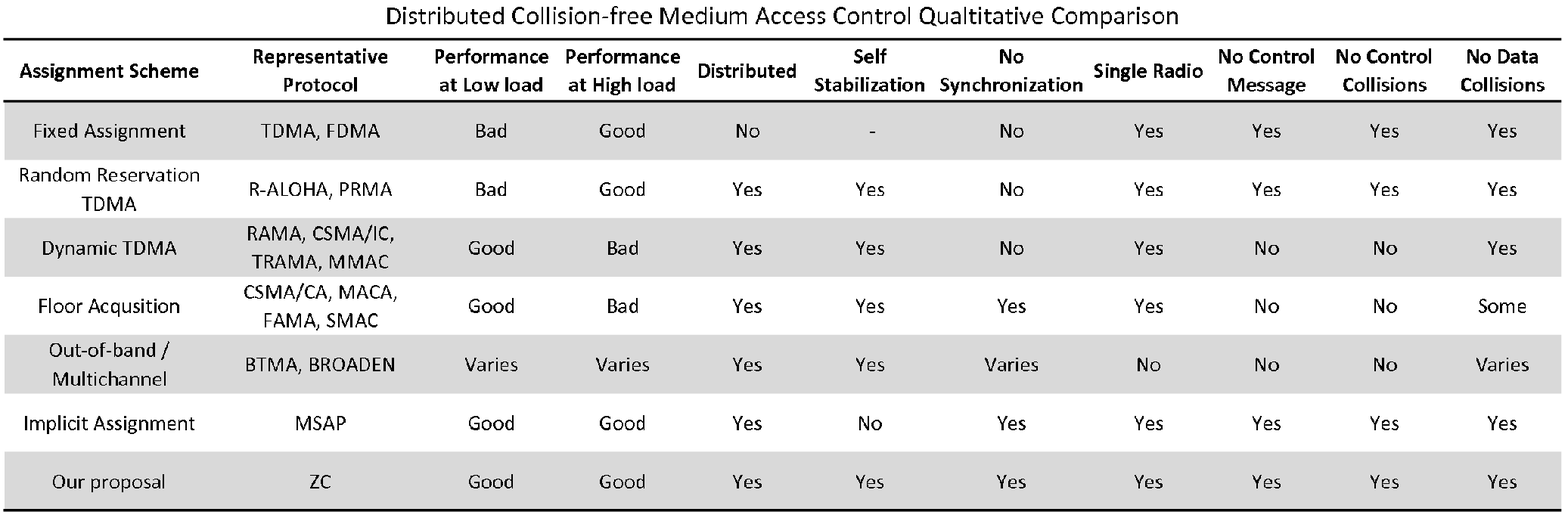, width=7.0in}
  \caption{Qualitative comparison of Distributed MAC protocols}
  \label{fig.comparison}
  \end{center}
\vspace{-0.2in}
\end{figure*}

A qualitative comparison of existing collision-free MAC protocols is presented in Fig. \ref{fig.comparison}.

\section{Related works}
Collision avoidance or collision-free distributed medium access control has a long history that we briefly review below. The majority of the protocols can be classified into one of following types: {\em reservation TDMA}, {\em dynamic TDMA}, {\em floor acquisition},  and {\em out-of-band or multichannel signalling}.


In a reservation or dynamic TDMA protocol, a device reserves some future epochs to transmit its packets. The random reservation TDMA protocols have many variations. R-ALOHA \cite{1973RALOHA}, PRMA \cite{1989PRMA} and their vast number of derivatives adopts ``reserve-on-success.'' There, time is divided into a sequence of frames and each frame consists of fixed-length slots. If a trial transmission is successful during a non-reserved slot within a frame, the corresponding slot of the following frames are regarded as reserved. \cite{2000FANTACCI} \cite{1983TASAKA} analyze performance and stability of this class of protocols.

Some protocols as in \cite{1983FALK} and \cite{1993WILSON} are classified as dynamic TDMA. Time is divided into a sequence of frames and each frame consists of at least two constant-length phases - one control phase which is used for competition for slot allocation via random access, auction \cite{1994RAMA} \cite{2002MICIC} \cite{2003CSMAIC}, or distributed election \cite{2003RAJENDRAN} and the remaining phase(s) which is (are) devoted to data \cite{1993WILSON}. Stations are synchronized at start of each phase, except for some cases as in \cite{2005MMAC} and \cite{1990ZHANG}. TRAMA \cite{2003RAJENDRAN}, MMAC \cite{2005MMAC} and their derivatives belong to this class in the context of wireless sensor networks.

While the slot is regarded as reserved in reservation TDMA until its reservation is revoked, the reservation is typically valid for a single frame time in dynamic TDMA. An inevitable disadvantage of TDMA frame based protocols, such as random reservation TDMA or dynamic TDMA, is that they exhibit poor utilization and excessive delays when only a few portion of stations in the network are active.

On the other hand, floor acquisition protocols are suitable for bursty traffic. Floor means exclusive channel access right, and is designed to relieve collisions from hidden stations. CSMA/CA \cite{1997IEEE80211}, MACA \cite{1990MACA}, MACAW \cite{1994MACAW} \cite{1994MACAW2}, FAMA \cite{1995FAMA}, CARMA \cite{1998CARMA} and their many descendants belong to this category. The main difference from dynamic TDMA is that transmission epoch needs to be immediately after the floor acquisition while in dynamic TDMA it is reserved for certain future epochs. Although these protocols enable use of statistical multiplexing, they require frequent control message exchange typically known as RTS/CTS dialogue overhead, which can be of substantial consumption in channel resource. Besides, control messages are not protected to be collision-free since they still follow random access. This scheme enables medium access in multiple collision domain at a cost of channel underutilization of exposed stations.
SMAC \cite{2004SMAC}, TMAC \cite{2003TMAC} and their many derivatives for energy limited wireless sensor networks belong to this class.

ZC is neither TDMA-based nor floor-acquisition based; it supports variable transmission time and no control dialogue exchange is required. ZC is asynchronous in the sense that it does not require explicit global / local synchronization. A ZC station is only required to count the elapsed time from the end of channel activity that it last observed. Because of these features, ZC exhibits good performance both at high and low traffic load.


There are protocols utilizing out-of-band signalling such as BTMA \cite{1975KLEINROCK2}, DBTMA \cite{2002DBTMA} or multichannel control such as SRMA \cite{1976KLEINROCK3} BROADEN \cite{2003BROADEN}. While those schemes enable distributed collision-free medium access, they demand multiple radios per device so it incurs complexity and cost. Therefore in our research, we focus on only the single radio case.

The closest relative of ZC is MSAP \cite{1980MSAP}. In MSAP, the sequence of mini slots plays a role analogous to a silent polling sequence  and the station with token does not release the channel until it empties its buffer. Moreover, in MSAP the access sequence is assumed to be predefined and shared by all stations ahead of time, and no distributed self-stabilizing process is proposed. So MSAP cannot recover from any dynamic event or carrier sensing error. In contrast, ZC is a simple but robust self-stabilizing, solving the time assignment problem in an effective way. BRAM \cite{1979BRAM} and SUPBRAM \cite{1985SUPBRAM} are extensions of MSAP with varying length of idle mini slots, but these protocols require packet decoding and global synchronization at every node.

There are also centralized algorithms to solve the assignment problem, but they frequently face feasibility challenges in collecting and redistributing information. As a result, such schemes are out of the topic of this research.

The remaining of the paper is structured as follows. Section \ref{s.algorithm} describes the operations of the protocol.  Section \ref{s.analysis} provides an analysis of the convergence time of the algorithm. Section \ref{s.experiment} shows the performance of the ZC protocol in a variety of metrics based on NS-2 simulations with realistic assumptions.  Section \ref{conclusion} summarizes the results of the paper and discusses future work.

\section{ZC Protocol} \label{s.algorithm}
We describe the operations of ZC and we illustrate them on a simple example.  We then comment on a key design assumption.

\subsection{Protocol Description}
When using the ZC protocol, in the steady state, the stations transmit in a round-robin order.  The total number of transmissions in one round is upper-bounded, but can be adapted later. In the basic operation, each station can transmit a single variable-length packet (typically $200\mu - 2200\mu$ seconds long) during one round (extension of this is explained shortly). If a station skips its turn, the medium is left idle for an empty $20\mu$ second-long mini-slot duration and the next station can access the medium after that. The length of a mini-slot is chosen to be substantially longer than a round-trip propagation across a few kilometer-wide network. To select its order of transmission, a station observes the sequence of idle mini-slots or busy transmission slots. Either one idle mini-slot or one transmission slot is called a virtual slot. Virtual slots correspond to transmission opportunities.  For instance, say that the number of transmissions per round is 8 and that a station observes the sequence $TETTEETE$ where $T$ designates a transmission of a variable-length packet and $E$ an empty mini-slot. In steady state, a station is able to count the sequence of virtual slots, and therefore find its own reservation.
In the transient state, when not all the stations have their own reservation yet, each station \textit{tries and sees} if a random choice of virtual slot works for itself. Suppose a station chooses the third empty mini-slot as its virtual slot after observing one round of medium access. If the station collides with another station that happened to select the same slot, the colliding stations repeat their random selection of an empty mini-slot in the next round.  Once a station has selected a successful order, it keeps it.  If a station stops transmitting for a given number of rounds, the station loses its reservation. The number of transmissions in a round can be adapted automatically to the number of active stations. Note that the time length of one round varies as the number of active stations and packet lengths vary.

Any collision caused by a new node arrival, carrier sensing error or clock drift triggers a new virtual slot selection process. This process is a robust self-stabilization mechanism: it converges to a zero collision state within at a few seconds, in the worst case for a reasonably big network and typical system parameters. After the convergence, the network operates without collision until another event triggers the selection process.

Due to its collision-free property and the relatively small size of mini-slots compared to transmission times, ZC achieves almost the maximum possible channel utilization, irrespective of the fraction of active stations. Using analysis and simulation we show that self-stabilization is fast enough to recover from collisions, channel randomness and carrier sensing error, and that ZC supports desirable coexistence of delay sensitive and delay tolerant traffic, which is not the case for WiFi networks.


On power up or arrival, station $C$ randomly and uniformly selects one of the unreserved virtual slots for its own transmission. (The case in which stations may select multiple slots will be explained later.) Since the station initially has no knowledge about the network, it is possible that the selected slot is in fact already reserved by another, implying potential collisions later. That station transmits during the selected virtual slot if it has a packet ready to transmit. If the transmission is successful, as the station detects when it gets an acknowledgement within half of a mini slot time from the end of its transmission, the station remembers that the slot is reserved for itself. Assume for now that there are neither random carrier sensing errors nor capture effects. Then the successful transmission within a virtual slot implies (a) there was no other transmitting station within the interference region in that slot, (b) all the other stations in the interference region sensed the slot to be busy if they were awake, and (c) they marked that slot as reserved and will not trespass in future as long as they already have not reserved it for themselves. Therefore the station is guaranteed to be collision-free in any future channel access as long as there is no other owner of the slot and a perturbation event does not occur.

If the transmission was not successful, the station gives up that slot and randomly and uniformly selects one of the unreserved virtual slots again. This process repeats until it finds a successful slot. Even though there are more than one owner of a virtual slot, as long as they do not experience a collision due to sparse traffic, they may enjoy statistical multiplexing over the same slot. Reservation is not permanent. If it were, the protocol could not be stabilized upon dynamic events of the network. If a virtual slot is sensed not to be used for a certain amount of time \rt, the slot becomes unreserved again.

If a carrier sensing error or clock drift occurs, it is possible that a station unintentionally trespass by mistaking other's reservation for its own one. If this happens, the colliding stations simply follow the above described protocol to resolve the issue: search again for an exclusive slot. It is important that this self-stabilization process is fast enough to deal with randomly occurring perturbations. Section \ref{s.analysis} shows the worst case bound of mean convergence time in analysis and simulation.

While ZC is designed for ad hoc mode operation, it provides more features under the infrastructure mode operation. If an access point(AP) exists, it may designate a slot as an anchor slot during which only beacon messages are broadcast. The anchor slot is not changed on collision because the AP does not rely on acknowledgement anyway. Then a power saving station can retrieve its previously reserved slot on wake up, without losing synchronization with the network and without incurring collisions. Secondly, the AP may reserved multiple slots since it typically requires multiple times of channel access to serve its associated stations. By allowing asymmetric assignment, the network performance can improve.

In order to complete the description, we discuss a few special cases.
It is possible that the transmission is still successful even though a collision occurs. Define $(\cdot, \cdot)$ be the distance metric. Consider two transmitters $A$ and $B$ and their receivers $a$ and $b$ respectively. Suppose $(A,a) << (A,B)$ and $(B,b) << (A,B)$. Then although $A$ and $B$ access the channel simultaneously, $a$ regards $A$'s transmission as signal and $B$'s as background noise, decoding $A$'s message successfully. $a$ will return acknowledgement. Similarly so does $b$. In this case ZC still properly functions. This spatial reuse phenomenon is confirmed via simulation. One interesting interpretation of this phenomenon is the sum of individual throughputs is larger than the naive channel capacity.

If there are more active stations than the network capacity, that is $M > N$, the collision-free channel access is not guaranteed any more. If all stations are backlogged, all virtual slots will be always used and some stations will face collisions while the rest will not. For those of colliding stations, since there is no unreserved slot, they will not move around and stick to their colliding slots. This makes the rest non-colliding stations not bothered and enables positive network throughput. It is not immediately clear that ZC in this case still outperforms CSMA. Simulations in \ref{s.experiment} show that ZC still outperforms CSMA and even the network throughput is lower bounded.
If stations are not backlogged and have independent packet arrival, some level of statistical multiplexing will be taken part in, together with ZC self-stabilization process on collisions.
Before network congestion, ZC, CSMA and TDMA does not show big difference in throughput and mean inter-access delay. Afterwards, performance varies depending on how sparse the traffic is and how big the network is. However, we do not study this issue in detail.


\subsection{Comments on Bounded Number of Stations}
ZC assumes that the number of active stations is bounded in order to guarantee its complete collision avoidance operation. This assumption can be justified because
\begin{itemize}
    \item   Physical systems have hard capacity limits because of implementation issues. As an example, the IEEE 802.11 MAC cannot support more than 2008 stations, which is the maximum length of the {Partial Virtual Bitmap} of {Traffic Indication Message} in {the Beacon frame}. Partial Virtual Bitmap is used to wake up power-saving stations.
    \item   Every network technology has its own coverage limit and it is unnatural to pack more than a certain number of stations into the coverage of a single network.
    For example, IEEE 802.3 100BASE-T and IEEE 802.3ab 1000BAST-T have a limited cable distance up to 100m. Also, the typical operating ranges of IEEE 802.11b, 802.11g and 802.11a are 100m, 50m and 20m respectively. Given the technology, the operator of the network already have decided the network capacity.
    \item   Per-user performance becomes unusable after a certain threshold of the network size. As the network size increases, severe performance degradation in terms of network throughput, per-user throughput, and transmission delay is induced. After a certain threshold of the network size, infinite capacity loses its meaning.
    \item   A wireless user's mobility is quasi-static. Think of a conference room in which stations sporadically arrive or depart. Inter-event(arrival or departure) time is on average expected to be more than tens of seconds at least. During this period, the network is static and we can exploit this feature.
\end{itemize}
Therefore in the following section, we analyze the convergence time to a zero collision state with a hard capacity limit in network size. In fact, if the network size is unbounded, convergence does not make sense.

\section{ZC Analysis}
\label{s.analysis}
In this section we show that ZC converges to a zero collision state in finite time. We then analyze the average convergence time and we derive a simple upper bound on that time. We conclude the section by showing that the convergence holds for an arbitrary sequence of wake up times of stations.

\subsection{Convergence}
It is fairly immediate to show that, for a single collision domain network with $M \leq N$,  the ZC algorithm is guaranteed to converge  to a zero collision state defined next.

\definition{Zero collision state}\\
We say that the network has reached a {\em zero collision state} when all $M$ stations have reserved a different slot.

\definition{Convergence time}\\
The {\em convergence time} of ZC is the first time when the network reaches a zero collision state.

\begin{theorem}{Convergence theorem}\\
Assume that \rt is finite and that \m $\leq N$. Then the ZC algorithm reaches a zero collision state in finite time.
\end{theorem}

\begin{proof}
Let $x_n$ be the number of stations with a reserved slot after $n$ idle slots.  If $x_n < M$, there is positive probability that one station will be alone in transmitting in the next idle slot and will consequently reserve a slot.  Consequently, $M$ is an absorbing state for the finite Markov chain $\{x_n, n \geq 0\}$.
\end{proof}

The following section derives an upper bound on the average convergence time. We then provide a simpler upper bound.

\subsection{Convergence Time Analysis}
It is important to show that the algorithm converges fast to a zero collision state.  Indeed, in practice, stations are always joining and leaving the network and the algorithm should quickly converge to a new zero-collision state for it to have a high throughput. In this section we study the convergence time of ZC and derive an upper bound on the mean convergence time. This upper bound assumes that a station attempts to transmit only once in $N$ consecutive slots instead of trying in each unreserved slot with some positive probability.  We consider the number $z_n$ of stations with reserved slots after the $n$-th cycle of $N$ slots.  It is clear that $\{z_n, n \geq 0\}$ is a Markov chain.

First, we derive the transition matrix of that Markov chain. Second, we calculate the average number of cycles until $z_n = M$. Third, we calculate the average total duration of those cycles.

\begin{theorem}{Probability of Reservation}\\
Consider a set of $M$ stations that select independently and uniformly one of $N$ slots.  The probability that exactly $k$ stations among $M$ select a slot that
no other station selects is given by
\be p_{N, M}(k) = \sum_{j=k}^{M}(-1)^{j-k}{M \choose j}{j \choose k}\frac{N!(N-j)^{M-j}}{(N-j)!N^M} \ee
for $0 \leq k \leq M$.
\end{theorem}
\begin{proof}
Let $I_{i}$ be the event \{Station $i$ selects a slot that no other station chooses\}. Fix a set $\Lambda$ with $|\Lambda| = j$. Then
\be
    P(\bigcap_{\lambda \in \Lambda}I_{\lambda})
    &=& {N \choose j}\frac{j!}{N^j}\left(\frac{N-j}{N}\right)^{M-j}\\
    &=& \frac{N!(N-j)^{M-j}}{(N-j)!N^M}.\sub
\ee
Also consider any set $\Gamma$ with $|\Gamma|=k$. Then
\be
    p_{N, M}(k)
    &=& \sum_{\Gamma : |\Gamma| = k } P(\bigcap_{\gamma \in \Gamma}I_{\gamma} \cap \bigcap_{\bar{\gamma} \in \Gamma^C }I_{\bar{\gamma}}^C) \\
    &=& {M \choose k} P(\bigcap_{\gamma \in \Gamma}I_{\gamma} \cap \bigcap_{\bar{\gamma} \in \Gamma^C }I_{\bar{\gamma}}^C).\sub
\ee
By the inclusion-exclusion principle,
\be
    P(\bigcap_{\gamma \in \Gamma}&I_{\gamma}& \cap \bigcap_{\bar{\gamma} \in \Gamma^C }I_{\bar{\gamma}}^C)\nonumber\\
    &=& \sum_{\Lambda: \Gamma \subset \Lambda} (-1)^{|\Lambda|-|\Gamma|} P(\bigcap_{\lambda \in \Lambda} I_{\lambda})\\
    &=& \sum_{j=k}^{M} \sum_{\Lambda: \Gamma \subset \Lambda, |\Lambda| = j} (-1)^{|\Lambda|-|\Gamma|} P(\bigcap_{\lambda \in \Lambda} I_{\lambda})\\
    &=& \sum_{j=k}^{M} (-1)^{j-k} {k \choose k} {M-k \choose j-k} P(\bigcap_{\lambda \in \Lambda} I_{\lambda}).\sub
\ee
Putting these expressions together produces the result.
\end{proof}

Note that
\[
p_{m, m + k} := P[z_{n+1} = m + k | z_n = m] = p_{M - m, N - m} (k)
\]
since, when $z_n = m$, there are $M - m$ stations left with unreserved slots and $N - m$ remaining slots to choose from.

Next we analyze the mean time until $z_n$ reaches the state $M$.  That is, let $L = \min \{n \geq 0 | z_n = M\}$.  We want to calculate $E[L|z_0 = 0]$.
Define $\beta (i) = E[L | z_0 = i]$.  Then one step equations are obtained and can be solved algebraically:

\be
\label{beta1}
\beta(i) &=&\left\{
\begin{array}{ll}
1 + \sum_{j=i}^{M}p_{i,j}\beta(j), & 0 \leq i \leq M-1\\
0, & i=M.
\end{array}\right.
\ee
Solving these equations yields $E[L | z_0 = 0] = \beta (0)$.
\begin{figure}
   \begin{center}
    \epsfig{file=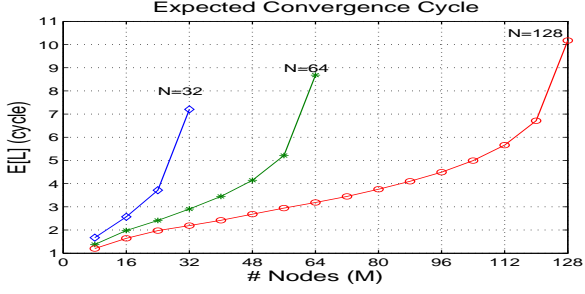, width=3.4in, height=1.5in}
   \caption{Expected Convergence Cycle}
   \label{fig.barking}
   \end{center}
\vspace{-0.2in}
\end{figure}

Finally, we derive the expected convergence time.

Let $\pi_n (k) := P[z_n = k]$, so that $\pi_n = \pi_0 P^n$ where $P$ is the transition matrix of the Markov chain $z_n$ that we derived earlier and $\pi_0 (m) = 1\{m = 0\}$.  Also, $P(L \leq n) = \pi_n (M)$.

Note that
\be
\pi_{n|L}(k) &:=& P(x_n = k|x_L = M) = \frac{\pi_n(k) \hat{\pi}^k_n(M)}{\pi_L(M)},
\ee
where $\hat{\pi}^k_n := \tilde{\pi}^k P^{L-n}$ and $\tilde{\pi}^k(j) := 1\{j=k\}$ for all $n \leq L$. The second equality comes from Bayes's rule.

Assume that cycle $n$ has $G_n$ successful transmissions, $V_n$ idle slots, and $B_n = N - G_n - V_n$ slots with collisions.  Designate by $t_g, t_v,$ and $t_b$ the duration of a successful transmission, of an idle slot, and of a collision, respectively. $t_s$ denotes the inter-slot gap. (See Appendix) Then
the length of cycle $n$, $T_n$, is given by
\be T_n &=& t_g G_n + t_v V_n + t_b B_n + t_s N \\
&=& (t_s + t_b)N + (t_v - t_b)V_n + (t_g - t_b)G_n . \sub\ee
Typically $t_b \approx t_g \gg t_v \geq t_s$ holds. Given the stopping period $L$, the total convergence time is $T = \sum_{n=1}^{L} T_n$. Therefore
\be \label{periodlength} E[T] &=& \sum_{l=1}^{\infty} \pi_l(M) \sum_{n=1}^{l} E[T_n|L=l],\ee
where
\[E[T_n|L] = (t_s + t_b)N + (t_g - t_b)E[G_n|L] + (t_v - t_b)E[V_n|L], \]
\[E[G_n|L] = \sum_{k=0}^{M} k \pi_{n|L}(k),\]
\[E[V_n|L] = \sum_{k=0}^{M} (N-k)(1-\frac{1}{N-k})^{M-k} \pi_{n-1|L}(k).\]\\
Putting the above expressions together yields the following result.

\begin{theorem}{Exact Expected Convergence Time}\\
ZC's convergence time can be computed as
\begin{eqnarray*}
E[T] &=& (t_s + t_b)N E[L]\\
&+& \sum_{l=1}^{\infty} \pi_{l}(M) \sum_{n=1}^{l} \sum_{k=0}^{M}\{ (t_g - t_b) k \pi_{n|l}(k) \\
&+& (t_v - t_b) (N-k)(1-\frac{1}{N-k})^{M-k} \pi_{n-1|l}(k)\}.
\end{eqnarray*}
\end{theorem}

\subsection{Simpler Upper Bound}
While exact and computable, the foregoing equation is less tractable and provides little physical sense. Instead, we derive a simpler upper bound.

In (\ref{periodlength}), $t_g - t_b$ can be either positive or negative. Since $0 \leq E[G_n|L] \leq M$, $N-M \leq E[V_n|L] \leq N$ and $(t_v - t_b) < 0$, if $(t_g - t_b)$ is positive, we find
\begin{eqnarray*}
E[T_n|L] &\leq& (t_s + t_b)N + (t_g - t_b)M + (t_v - t_b)(N-M) \\
&=& (t_s + t_v) N + (t_g - t_v)M\sub
\end{eqnarray*}
or if $(t_g - t_b)$ is negative,
\begin{eqnarray*}
E[T_n|L] &\leq& (t_s + t_b)N + (t_g - t_b)\cdot 0 + (t_v - t_b)(N-M) \\
&=& (t_s + t_v) N + (t_b - t_v)M.\sub
\end{eqnarray*}
Therefore,
\begin{theorem}{Upperbound of Expected Convergence Time}\\
\label{thm.upperbound}
When stations are backlogged and powered up synchronously, the expected convergence time is bounded by
\be \label{eqn.bound} E[T] &\leq& \left\{(t_s + t_v) N + (\max(t_g, t_b) - t_v) M \right\} E[L] \ee
where $E[L]$ is computed from (\ref{beta1}) and portrayed in Fig. \ref{fig.barking}.
\end{theorem}
\begin{figure}
   \begin{center}
   \epsfig{file=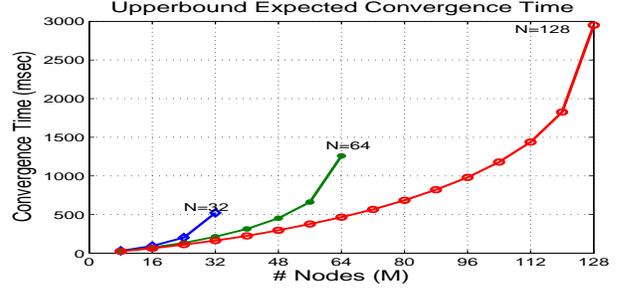, width=3.4in, height=1.5in}
   \end{center}
   \begin{center}
   \caption{Upper bound of Expected Convergence Time}
   \label{fig.upperbound}
   \end{center}
\vspace{-0.2in}
\end{figure}
The foregoing equation provides us an upper bound of the expectation of the convergence time. Recalling that $t_g$ is of order of a few milliseconds, that $N$ or $M$ are typically less than a few hundreds, and that $E[L]$ is around $10$ (Fig. \ref{fig.barking}), $E[T]$ is roughly less than a few seconds. Indeed, Fig. \ref{fig.upperbound} shows the upper bound of expected convergence time in IEEE 802.11b DSSS Long preamble MAC/PHY network with parameters $t_g = 2150 \mu sec, t_v = 20 \mu sec, t_b = 2266 \mu sec$, and $t_s = 0 \mu sec$.

For example, for the network of $N=128, M=128$, the upper bound of expectation of convergence time is $2.92sec$ which is notably close to the simulation result.

\subsection{More general cases}
Theorem \ref{thm.upperbound} holds in more general cases. So far we have considered the case where all the stations are powered up simultaneously. In a more typical case, stations join the network and become active in an asynchronous manner. In that case, when a later station arrives, a positive expected number of stations have already acquired their own access slot. By making a station scan the channel for one period before access, a station that arrives later competes only with the stations that have not yet reserved a slot. Therefore, the expected convergence time after the station arrives is always less than that of the synchronous arrival case.

We have developed and discussed the convergence time of the ZC algorithm under a saturated queue traffic model in which stations always transmit packets whenever they are allowed to do so, which is no realistic.  Suppose again that all the stations join the network at the same time. Due to the intermittent traffic generation, some stations skip their access chances, effectively reducing the number of `active' stations for the same network capacity $N$. Those stations that do not transmit during the initial converging period are automatically given additional scanning periods, which results in a reduction of the subsequent convergence time as discussed in the previous paragraph. 

\begin{figure}
   \begin{center}
   \epsfig{file=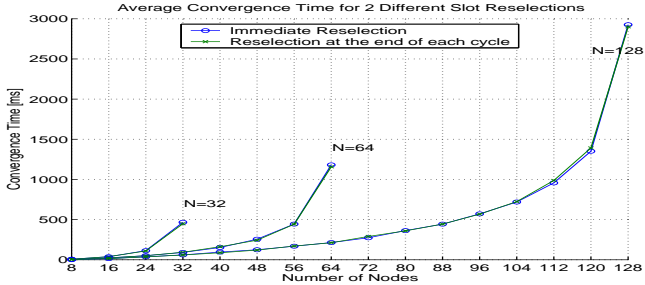, width=3.4in, height=1.5in}
   \end{center}
   \begin{center}
   \caption{Convergence Time of Immediate Reselection and Reselection at the end of each cycle}
   \label{fig.reselection}
   \end{center}
\vspace{-0.2in}
\end{figure}

In the preceding analysis, we implicitly assumed that slot reselection after collision is done at the end of every cycle. In practice, slot reselection is done immediately after collision. While these two schemes are obviously different, they effectively show the same convergence time in simulation (Fig. \ref{fig.reselection}).

\section{Proof of Concepts and Experimental Results}
\label{s.experiment}
\subsection{Proof of Concepts}
To verify the viability of the ZC algorithm, we implemented an NS-2 ZC simulator on top of the IEEE 802.11 PHY and MAC modules.
The modules support propagation delay, two-way ground channel model, signal-to-noise ratio computation and thresholding, capture effect, carrier-sensing
and collisions. The NS-2 ZC simulator modifies the behavior of the access time decisions and replaces the exponential random backoff.
To maximize the backward compatibility, the other features of IEEE 802.11 MAC are still used. Various types of traffic models are used in the simulations:
saturated queue traffic, VoIP Constant Bit Rate traffic, and web traffic on top of full TCP. Except for the result of convergence time and network goodput, the base station reserves $M-1$ slots when it serves $M-1$ stations. RTS/CTS was not used. In the remainder of the paper, CSMA refers to IEEE 802.11b WiFi.
\begin{table}[ht]
\small
    \centering \caption{Experimental configuration}
        \begin{tabular}{|c||c|c|c|}
        \hline
        {\bf{Parameter}}&{\bf{Notation}}&{\bf{ZC}}   &{\bf{CSMA}}\\
        \hline
        \multirow{2}{*}{PHY protocol}& &\multicolumn{2}{|c|}{IEEE 802.11b/DSSS} \\
        &&\multicolumn{2}{|c|}{Long Preamble} \\
        \hline
        Data TX rate &$R_{TX}$&\multicolumn{2}{|c|}{11 Mbps}\\
        \hline
        Slot time    &$T_{SLOT}$&\multicolumn{2}{|c|}{20 $\mu{}$sec}\\
        \hline
        SIFS&$T_{SIFS}$&\multicolumn{2}{|c|}{10 $\mu{}$sec}\\
        \hline
        DIFS&$T_{DIFS}$&\multicolumn{2}{|c|}{50 $\mu{}$sec}\\
        \hline
        Contention&\multirow{2}{*}{$N$}&Fixed 128&Dynamic\\
        Window&& (see detail)& 32-1024\\
        \hline
        Network size&\m&\multicolumn{2}{|c|}{4 - 128}\\
        \hline
        Traffic model    &&\multicolumn{2}{|c|}{Saturated queue/VoIP/Web}\\
        \hline
        Frame size&$x$&\multicolumn{2}{|c|}{2346 or varying bytes}\\
        \hline
        ACK size&$L_{ACK}$&\multicolumn{2}{|c|}{14 bytes}\\
        \hline
        PHY Preamble&$L_{Pre}$ &\multicolumn{2}{|c|}{18 bytes}\\
        PHY PLCP&$L_{PLCP}$&\multicolumn{2}{|c|}{6 bytes}\\
        \hline
        PHY TX rate&$R_{PHY}$&\multicolumn{2}{|c|}{1 Mbps}\\
        \hline
        Recycle Timer    &\rt&10         &0\\
        \hline
        \end{tabular}
      \label{table.expconfig}
\end{table}
\subsection{Convergence time}
\begin{figure}[ht]
  \begin{center}
  \epsfig{file=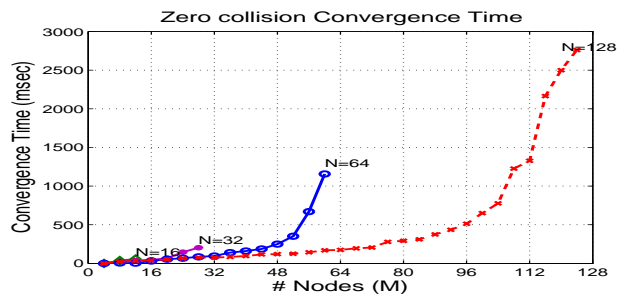, width=3.4in, height=1.5in}
  \caption{Convergence time}\label{fig.convergencetime}
  \end{center}
\vspace{-0.2in}
\end{figure}

Convergence time is defined as the time to reach a zero collision state from the moment of perturbation. The largest perturbation
is induced when all the stations have no history of the network and are powered up simultaneously. This leads to the worst case
of convergence time. Using backlogged traffic with large packet size (2346 bytes), Fig. \ref{fig.convergencetime} shows the convergence
time for a network with capacity 16, 32, 64 and 128 respectively. The largest network (128 nodes) reaches a zero collision state
within 3 seconds. After the convergence, the network experiences no collision and network performance including throughput and delay is
enhanced compared to the standard 802.11 protocols. Obviously, if $M > N$, the algorithm does not converge.
\subsection{Network Goodput}
\begin{figure}[ht]
    \begin{center}
    \epsfig{file=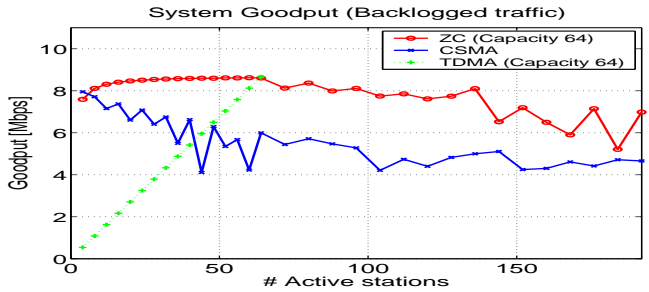, width=3.4in, height=1.5in}
    \caption{Network Goodput (Backlogged traffic)} \label{fig.backloggedgoodput}
    \end{center}
\end{figure}

\begin{figure}[ht]
    \begin{center}
    \epsfig{file=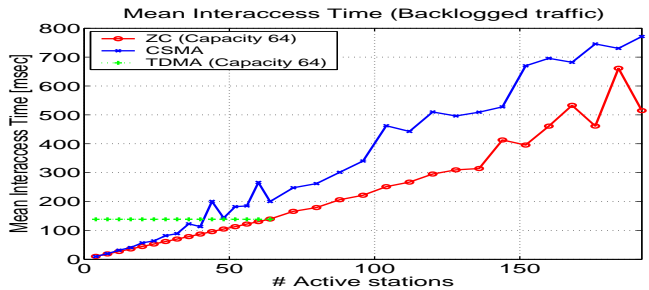, width=3.4in, height=1.5in}
    \caption{Mean interaccess Delay (Backlogged traffic)} \label{fig.backloggeddelay}
    \end{center}
\end{figure}

\begin{figure}[ht]
    \begin{center}
    \epsfig{file=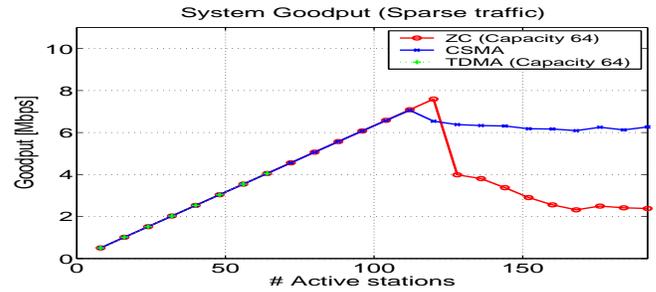, width=3.4in, height=1.5in}
    \caption{Network Goodput (Sparse traffic)} \label{fig.sparsegoodput}
    \end{center}
\vspace{-0.2in}
\end{figure}

Fig. \ref{fig.backloggedgoodput} and \ref{fig.backloggeddelay} show the fundamental difference of ZC, CSMA, and TDMA in terms of network goodput and mean interaccess time when the stations are backlogged. ZC and TDMA's network capacity is set to 64. For a fair baseline comparison, each station is allowed to take a single access chance in a cycle. Several observations are noteworthy: a) The fundamental performance difference of ZC from CSMA is ZC's goodput actually improves as the network size grows until it reaches the capacity and afterwards it starts to decrease. b) For a substantially large range of network size (6 - 192+), ZC outperforms CSMA even when $M > N$. c) While both ZC and TDMA do not experience collisions when $M \le N$, ZC always outperforms TDMA. Especially when the network size is small, ZC's performance is remarkably better. d) When the network size is very small ($M < 6$), ZC's mini slot size is not negligible any more and plays a role in contributing to performance degradation. This happens when the network capacity is too overestimated than the actual network size.
When the stations are not backlogged, `reserving the same slot' does not necessarily mean a collision. Indeed, stations can enjoy statistical multiplexing without generating collisions. In Fig. \ref{fig.sparsegoodput}, a station generates a 2346 Byte packet at every 300 msec. Although the network capacity is set to $N=64$, stations do not experience collisions almost until $M \approx 2N$. In that range ZC, CSMA, and TDMA show virtually identical goodput except that TDMA is not defined for $M > N$. After a certain threshold of the network size (here $2N$), collision effects are more pronounced in ZC than CSMA.
\subsection{Carrier sensing error effect}
\begin{figure}[ht]
    \begin{center}
    \epsfig{file=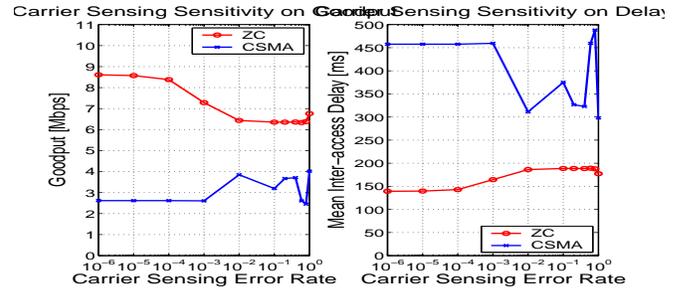, width=3.4in, height=1.5in}
    \caption{Carrier sensing and Clock drift error effect} \label{fig.clockdrift}
    \end{center}
\vspace{-0.2in}
\end{figure}
By design, ZC is sensitive to carrier sensing and asynchronous clock. Thus it is important to verify whether ZC's good performance shown in the previous section can be robustly maintained upon errors or clock drifts. The error model is as follows: each station performs carrier sensing and undergoes errors independently. An idle mini slot can be sensed to be busy with probability $p1$. Also an idle mini slot can be sensed to be idle but counted two times with probability $p2$ due to slight timing mismatch. A busy virtual slot can be sensed to be idle as well with probability $p3$, but it is unlikely to detect the whole portion of a busy virtual slot, which is in fact long in time, is sensed to be idle in the spatial range of consideration. Therefore in fact $p1$ and $p2$ provides enough perturbation for simulation purpose and we consider $p1=p2=p$ for simplicity and vary $p$ from $10^{-6}$ to $1$. Network size and capacity is fixed to 64 and all stations are backlogged. Results are shown in \ref{fig.clockdrift}.

As expected, CSMA is relatively insensitive to carrier sensing and clock drift errors while ZC is not.
However interestingly enough, irrespective of the error rate, ZC's goodput is lower bounded and does not drop to zero. Considering that ZC never achieves zero collision state, this is an unexpected result. Intuitively, more perturbation leads to more collisions. Since the number of virtual slots and active stations are fixed, more collisions imply more stations are concentrated in certain virtual slots. Then there are more non-conflicting slots, which contributes to goodput.

\begin{figure}[ht]
    \begin{center}\small
    \epsfig{file=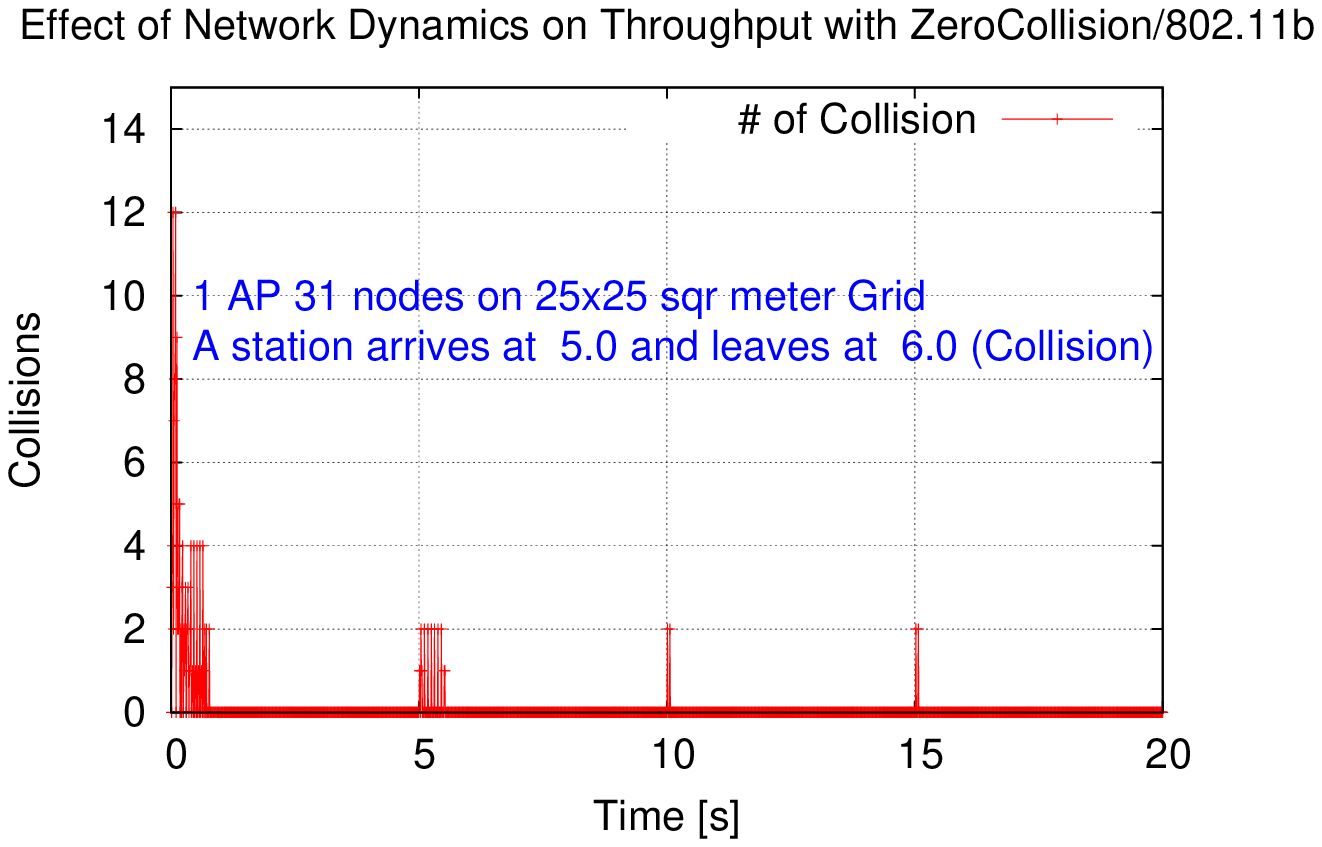, width=3.4in,height=1.5in}
    \normalsize
    \caption{Collisions on station arrivals}\label{fig.delay}
    \end{center}
\vspace{-0.2in}
\end{figure}

Fig. \ref{fig.delay} portrays the system dynamics when a station frequently comes and goes. Initially there are 31 stations and one AP, randomly located within an area of $25 \times 25$ square meters, with backlogged traffic and the stations are powered up simultaneously. After the initial convergence, the network achieves the maximal goodput. At 5, 10, 15 seconds in the simulation, a 32th station comes to the network with no prior information on the network, access the channel for 1 second with backlogged traffic, and then leaves. 
We can see that although a new arrival of a station momentarily affects the performance of the network, its collisions are quickly resolved and the network converges to a zero collision state.
\subsection{Performance for Delay sensitive and Delay tolerant traffic}
\label{s.voipdelaycapacity}
\begin{figure}[ht]
    \begin{center}\small
    \epsfig{file=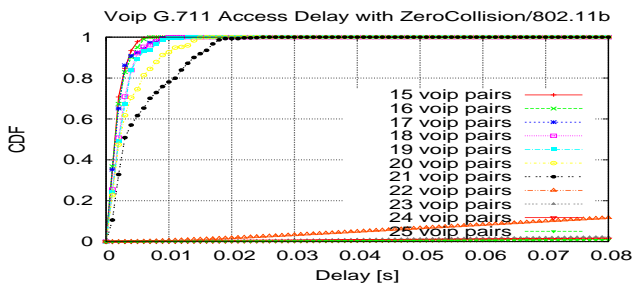, width=3.4in, height=1.5in} \\ (a) ZC \\
    \epsfig{file=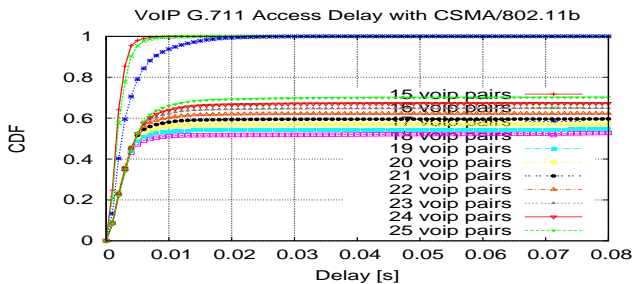, width=3.4in,height=1.5in}\\ (b) CSMA
    \normalsize
    \caption{G.711 VoIP Delay}\label{fig.voipdelay}
    \end{center}
\vspace{-0.2in}
\end{figure}
We compare the cumulative distributions of VoIP access delay for ZC/IEEE 802.11b and CSMA/IEEE 802.11b. Using a G.711 codec, each source of a VoIP conversation pair generates a 240-byte long application packet every 30msec. We assume that the performance is acceptable if 99th percent of the packets experience an access delay less than or equal to 30msec. The simulation shows that ZC support up to 21 conversation pairs while CSMA supports about 17.  Indeed by simple computation, it is easily shown that 21 conversation pairs actually fill up the 30msec time frame and are the maximum number of VoIP capacity with stable queue size and delay.

\begin{figure}[ht]
    \begin{center}\small
    \epsfig{file=voip-delay-802_11b.eps, width=3.4in, height=1.5in} \\ (a) ZC \\
    \epsfig{file=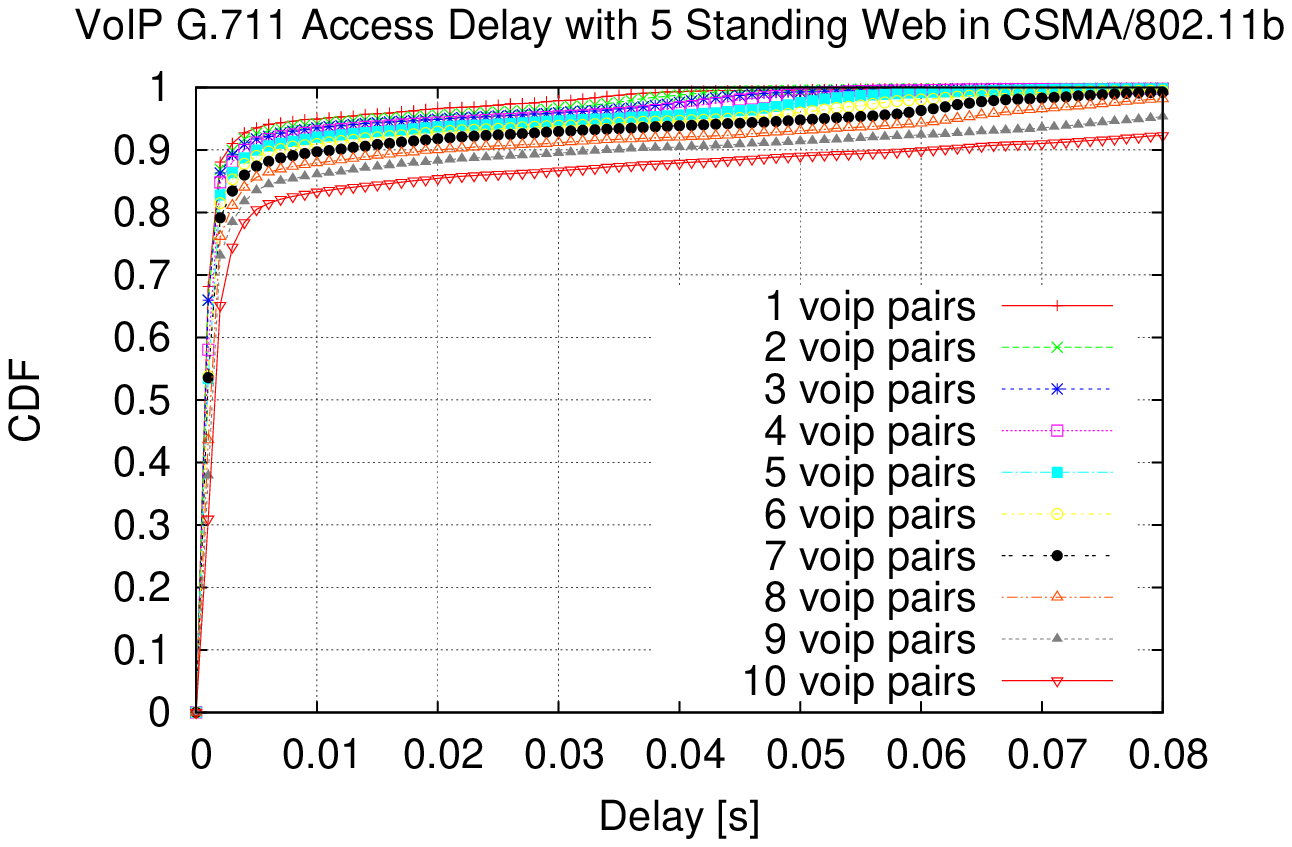, width=3.4in,height=1.5in}\\ (b) CSMA
    \normalsize
    \caption{G.711 VoIP Delay with 5 Standing Web }\label{fig.voipwebdelay}
    \end{center}
\vspace{-0.2in}
\end{figure}

Additional experiment is found more interesting in case of the coexistence with delay tolerant traffic. ZC is shown to support 18 times more VoIP sessions than CSMA when 5 background web sessions(\cite{2004PACKMIME} model) are ongoing. It has been known that the VoIP capacity of CSMA networks is significantly impaired by the existence of a few TCP flow. That phenomenon can be observed in Fig. \ref{fig.voipwebdelay}(b) where one VoIP session is barely supported. Different from CSMA, delay sensitive traffic and delay tolerant traffic mingle easily. With the same background Web traffic, the ZC network can support about 18 VoIP connections.
\begin{table}[ht]
\small
    \centering \caption{VoIP Capacity in IEEE 802.11b}
        \begin{tabular}{|c|c|c|}
        \hline
        {\bf{Traffic}}&{\bf{ZC}}&{\bf{CSMA}}\\
        \hline
        {VoIP only}&{21}&{17}\\
        {VoIP + 5 Web}&{18}&{1}\\
        \hline
        \end{tabular}
      \label{table.voipcapacity}
\end{table}
We also have web transaction delay distribution result under both ZC and CSMA. With 25 simultaneous web users and nominal simulation parameters, 90\% delay is achieved within 5 seconds under CSMA, while ZC achieves it within 2 seconds. To save the space, we omit the experimental CDF.


\subsection{Multiple Collision Domain Goodput}
\label{s.mc}
We have seen ZC outperforms both TDMA and CSMA in a single collision domain in the foregoing sections. In a multiple collision domain, neither ZC, CSMA nor self-regularizing distributed TDMA is collision-free. While ZC is originally designed for the operation in a single collision domain, it does not mean it is not operable in a multiple collision domain environment. Rather, simulations in complicated multiple collision domain topologies show the evidence that ZC is commensurate with CSMA in terms of goodput.
To emulate a complex wireless sensor network environment with signal-blocking walls and buildings, we generate random topologies in a following manner: in a $200 \times 200$ sq. meter area, we place $N$ nodes at random location. There are $\frac{N}{2}$ distinct and independent flow pairs, each of which has approximately $600Kbps$ source rate. Each node has a random connectivity with another with probability $\gamma$ irrespective of its geographical location. Here the connectivity between two nodes means they can hear each other's transmission. No connectivity implies there is a wall so that two nodes cannot hear each other at all. A flow pair is always set to have a connectivity. On average, each node has connectivities with $\gamma$ fraction of $N$ nodes independently from others' connectivity state, and it is hidden from $1-\gamma$ fraction of $N$ nodes. Connectivity is assumed to be reciprocal.

\begin{figure}[ht]
    \begin{center}\small
    \epsfig{file=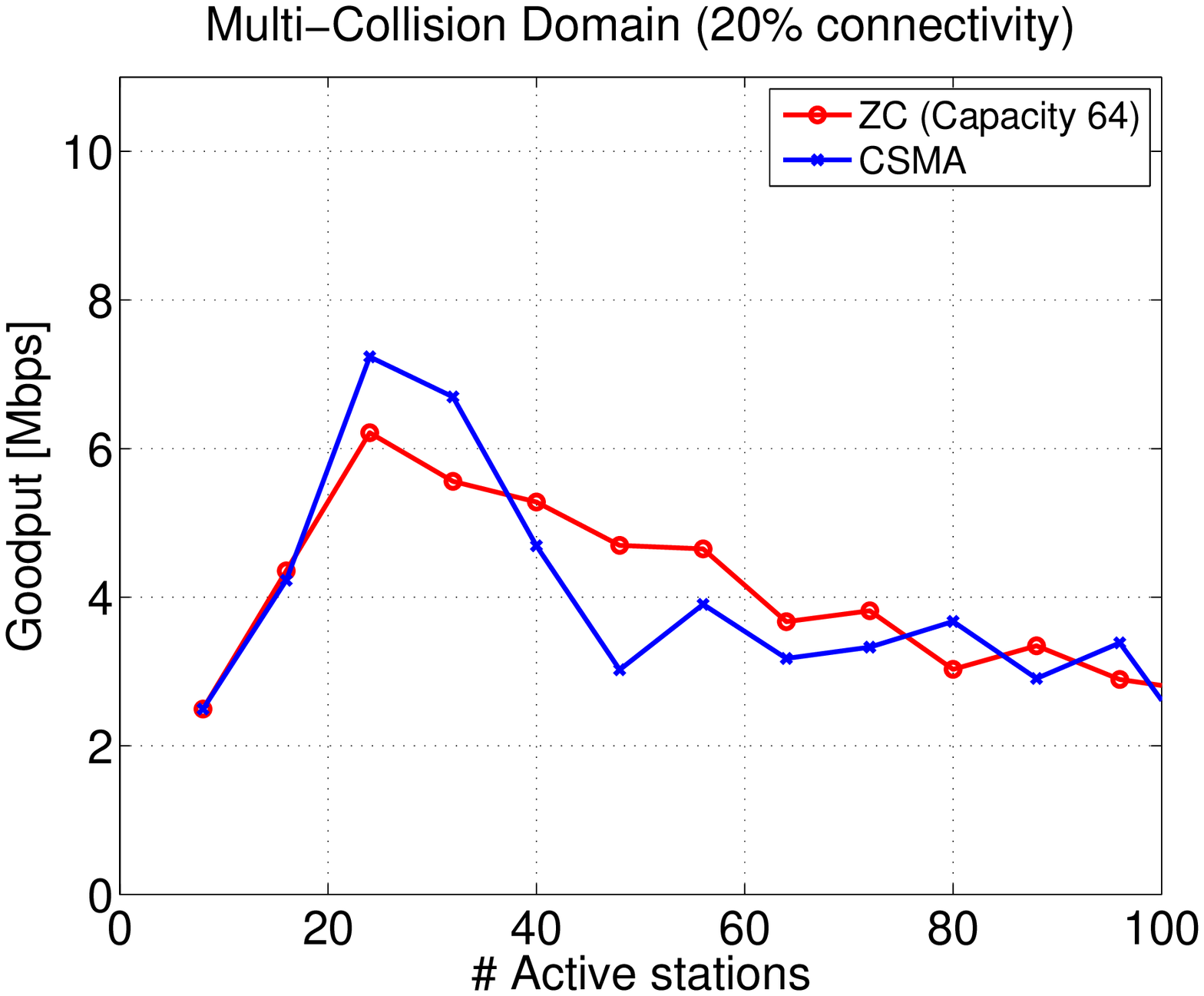, width=3.4in, height=1.5in} \\ (a) Connectivity $\gamma=0.2$ \\
    \epsfig{file=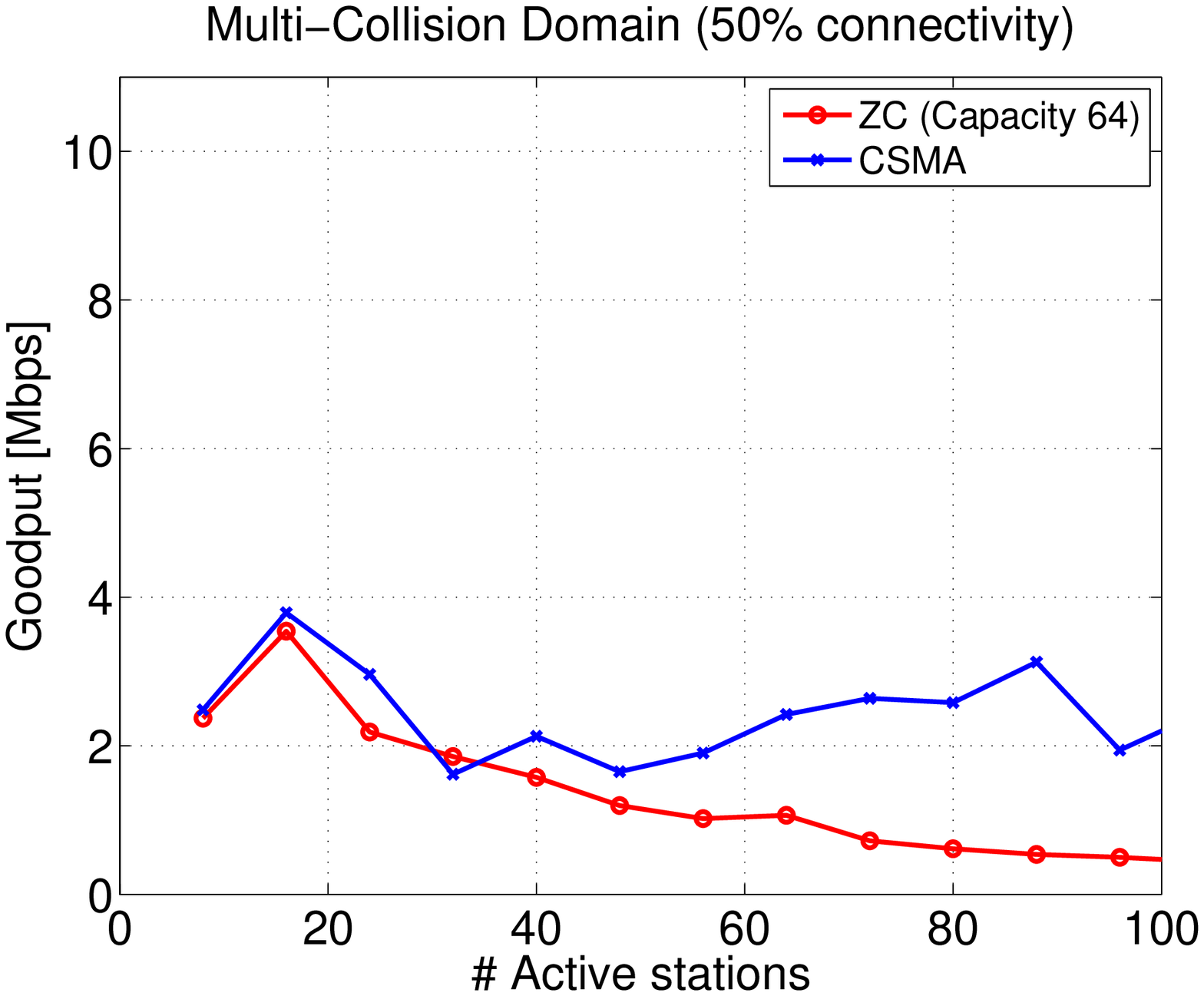, width=3.4in,height=1.5in}\\ (b) Connectivity $\gamma=0.5$
    \normalsize
    \caption{Multiple collision domain goodput comparison: ZC and CSMA}\label{fig.mc}
    \end{center}
\vspace{-0.2in}
\end{figure}

Obviously the goodput degradation is mainly caused by hidden node/exposed node syndrome in the multiple collision domain. Experimental results show when the fraction of random connectivity is low with, say, $\gamma \le 0.2$, ZC is comparable to CSMA. When $\gamma=0.5$, note that this topology is excessively complex, and carrier sensing based medium access is more error-prone. The more the MAC relies on carrier-sensing, the more collisions are likely to occur, which is the case of ZC.

\section{Conclusion}
\label{conclusion}
We proposed a zero collision achieving asynchronous distributed medium access control, called ZC, which provides superior performance compared to CSMA and TDMA in terms of goodput and mean interaccess delay. By design, ZC is sensitive to network dynamics, carrier sensing error, or corresponding clock drift. Analytically we showed that ZC's mean convergence time to zero collision state is upperbounded and 2-3 seconds in worst case for a reasonably big network size. Empirical results show that even at a severe carrier sensing error rate, ZC robustly maintains superior performance to CSMA.
ZC can be easily implemented using 802.11 hardware following the same PHY and most of MAC specification. Although all the performance figures are based on IEEE 802.11 PHY families in this paper, its application can be easily extended to other well-known wireless or wired technologies.


%
\bibliographystyle{abbrv}
\bibliography{ZeroCollision}  
%
%

\appendix
\section{Definitions of parameters}
Based on IEEE 802.11 MAC/PHY specification, timing parameters used in the previous formula is defined as follows:
\be
t_g &=& 2 \times (T_{PLCP_{Preamble}} + T_{PLCP_{Header}})\\
    &+& T_{MPDU} + T_{SIFS} + T_{Ack}\\
t_b &=& T_{PLCP_{Preamble}} + T_{PLCP_{Header}} \\
    &+& T_{MPDU} + T_{EIFS}\\
t_v &=& T_{SLOT}\\
t_s &=& 0\sub
\ee
\balancecolumns
\end{document}